\documentclass[aps,prd,showpacs,superscriptaddress,groupedaddress]{revtex4}  
\usepackage{graphicx}  
\usepackage{dcolumn}   
\usepackage{bm}        

\usepackage{amsfonts}

\usepackage{amssymb,latexsym,amssymb,amsmath,amsbsy,amsopn,amstext,upgreek}
\usepackage{color}
\usepackage{graphicx,wrapfig,fancybox,watermark,graphics}
\usepackage{pgf}
\usepackage{float}

\newcommand{\figdir}[1]{figures/#1}

\def\rms{root mean square~}
\newcommand{\figref}[1]{Figure (\ref{#1})}

\newcommand{\secref}[1]{Section \ref{#1}}
\newcommand{\healpix}{HEALpix~}
\newcommand{\nside}[1]{N_{\rm side}}

\begin{document}

\title{Cosmological Applications of the Gaussian Kinematic Formula\thanks{%
Research supported by ERC Grant 277742 Pascal.}}
\author{Yabebal Fantaye} 
\affiliation{Dipartimento di Matematica, Università di Roma''Tor Vergata,'' Via della Ricerca Scientifica 1, I-00133 Roma, Italy}
\author{Frode Hansen}
\affiliation{Institute of Theoretical Astrophysics, University of Oslo, PO Box 1029 Blindern, 0315 Oslo, Norway }
\author{Davide Maino}
\affiliation{Dipartimento di Fisica, Universit\'{a} di Milano, Via Celoria 16, I-20133, Milano, Italy}
\author{Domenico Marinucci}
\affiliation{Dipartimento di Matematica, Università di Roma''Tor Vergata,'' Via della Ricerca Scientifica 1, I-00133 Roma, Italy}


\date{\today}

\begin{abstract}
The Gaussian Kinematic Formula (GKF, see Adler and Taylor
(2007,2011)) is an extremely powerful tool allowing for explicit
analytic predictions of expected values of Minkowski functionals
under realistic experimental conditions for cosmological data
collections. In this paper, we implement Minkowski functionals on
multipoles and needlet components of CMB fields, thus allowing a
better control of cosmic variance and extraction of information on
both harmonic and real domains; we then exploit the GKF to provide
their expected values on spherical maps, in the presence of
arbitrary sky masks, and under nonGaussian circumstances. All our
results are validated by numerical experiments, which show a
perfect agreement between theoretical predictions and Monte Carlo
simulations.
\end{abstract}




\maketitle


\section{Introduction}

A general trend in modern cosmological research is the implementation of
more and more sophisticated statistical tools to perform data analysis.
Indeed, as well-known cosmological data have reached over the last decade an
unprecedented accuracy, so that it has become customary to speak about a
golden era for Cosmology, featuring a data deluge from a bunch of satellite
- and ground based-experiments. As the data grow in size and precision, more
and more detailed questions can be addressed, and exploiting techniques at
the frontier of statistical and mathematical research becomes mandatory to
warrant a full exploration of the available evidence.

Among these techniques, stochastic geometry tools have now become
very well established, especially in the field of Cosmic Microwave
Background radiation experiments. In this area, one of the most
popular geometric tools for data analysis are certainly the
so-called Minkowski functionals (MFs), which have been extensively
exploited as tools to search for nonGaussianities, anisotropies,
asymmetries and other features of CMB data.  The use of MFs in
Cosmology goes back at least to \cite{tomita1986_genus,
Coles1988_mfCMB}; a complete bibliography would certainly include
hundreds of entries, so we refer only to the earlier works by
\cite{schmalzing1997_mf,
  dolgov1999_mfpolm, naselsky1998_mfPol, Novikov2000_mfCMB,
  schmalzing1998_mfreview,matsubara2003_mf2PT,
  wmap2003_wmf,wmap2007_wmf} and to the more recent
ones by \cite{natoli2010mfBoomrang, matsubara2010_mfFnl,
  ducout2013_mfFnl, gratten2012_mfLSSreview, munshi2013_mfSkewCl,
  planck2013_IS}.

As well-known, on the plane there are three Minkowski functionals \emph{M}$%
_{0},$\emph{M}$_{1},$\emph{M}$_{2}$ which can be taken to
represent, respectively, the area, the boundary length and the
Euler-Poincar\'{e} characteristic (number of connected components
minus holes) of any given region. To characterize the behaviour of
data from a random field ($T(x),$ say) it is has then become
customary to consider flat-sky approximations and
to focus on the excursion sets%
\[
A_{u}(T):=\left\{ x:T(x)>u\right\} \text{ ,}
\]%
e.g. the regions of the plane where the value of $T$ exceeds the threshold $%
u;$ the corresponding functionals \emph{M}$_{i}(A_{u}(T)),$
$i=0,1,2$, can then be computed for real data with a number of
accurate and numerically efficient packages. The expected values
of the Minkowski functionals in the planar case and under
Gaussianity is analytically known to the literature since the work
of Adler in the early 80's (\cite{adler1981}, see also
\cite{tomita1986_genus}), and these predictions can be compared to
values on observed data to implement a number of statistical tests
(see for instance \cite{planck2013_IS} and the references
therein).

In the last decade, some major progresses have occurred in the
mathematical understanding of the geometry of random fields,
namely the discovery of the Gaussian Kinematic Formula by Taylor
and Adler (see \cite {TaylorAdler2003, Taylor2006,
TaylorAdler2009}, \cite{adlerstflour}, \cite{RFG}).

As we shall discuss in the next section, the Gaussian kinematic
formula allows a simple computation of the expected values for
Lipschitz-Killing curvatures (equivalent to Minkowski functionals,
see below) under an impressive variety of extremely different
circumstances, covering arbitrary manifolds with and without
masked regions and a broad class of nonGaussian models. These
expected values take extremely neat and intuitive forms, and can
be immediately compared to simulations and observed data. One of
our purposes in this paper is to exploit these recent results to
develop a number of analytic predictions on functionals tailored
to test nonGaussianities and asymmetries on CMB data.

More precisely, in this paper we aim at the implementation of Minkowski
functionals/Lipschitz-Killing curvatures on the multipole and needlet
components of observed data. To be more explicit, we start from the
decomposition of an observed spherical (e.g., CMB) map into harmonics as
\begin{equation}
T(x)=\sum_{\ell =1}^{L_{\max }}\sum_{m=-\ell }^{\ell }a_{\ell m}Y_{\ell
m}(x)=\sum_{\ell =1}^{L_{\max }}T_{\ell }(x)\text{ ;}  \label{specrap}
\end{equation}%
It is well-known that the decomposition \eqref{specrap} is only feasible for
unmasked (full-sky) data, a condition which is usually considered very
difficult to meet for CMB experiments (see, however, the recent full-sky
maps produced by \cite{starketal2014}). To handle masked regions, it has
hence become very popular to introduce various forms of spherical wavelets,
which enjoy much better localization properties than spherical harmonics in
the real domain, and are therefore much less affected by sky cuts. In this
paper, we shall focus in particular on the needlet system, which is defined
by the filter%
\[
\psi _{jk}(x)=\sum_{\ell ,m}b(\frac{\ell }{B^{j}})\overline{Y}_{\ell m}(\xi
_{jk})Y_{\ell m}(x)\text{ ,}
\]%
where $\left\{ \xi _{jk}\right\} $ denotes a grid of points on the
sphere (such as HealPix centers at a given resolution, see
\cite{healpix}), $B>1$ is some fixed bandwidth parameter and the
weight function $b(\frac{\ell }{2^{j}})$ satisfies three
conditions, namely a) it is compactly supported in the interval
($B^{-1},B);$ b) it is smooth; c) the partition of unity property
holds, e.g. $\sum_{j}b^{2}(\frac{\ell }{B^{j}})=1$ for all $\ell
.$ Needlets have been shown to enjoy very good localization
properties in the real domain; needlet coefficients are given by
the projection
\[
\beta _{jk}=\int_{S^{2}}T(x)\psi _{jk}(x)dx=\sum_{\ell
=B^{j-1}}^{B^{j+1}}\sum_{m}b(\frac{\ell }{B^{j}})a_{\ell m}Y_{\ell m}(\xi
_{jk})\text{ ,}
\]
and they allow for the reconstruction formula
\begin{eqnarray}
T(x) &=&\sum_{j=1}^{J_{\max }}\sum_{k}\beta _{jk}\psi
_{jk}(x)=\sum_{j=1}^{J_{\max }}\beta _{j}(x)\text{ , }  \label{needrap} \\
\beta _{j}(x) &=&\sum_{k}\beta _{jk}\psi _{jk}(x)=\sum_{\ell
=B^{j-1}}^{B^{j+1}}\sum_{m}b^{2}(\frac{\ell }{B^{j}})a_{\ell m}Y_{\ell m}(x)
\text{ ,}  \nonumber
\end{eqnarray}%
see \cite{npw1}, \cite{bkmpAoS}, \cite{mpbb08},
\cite{donzelli2012_mfNeedlet} for further discussions and applications to some CMB data
analysis issues.

Our aim is to apply Minkowski functionals on both the field
components $\left\{ T_{\ell }(x),\beta _{j}(x)\right\} $ rather
than on the original map. This form of harmonic/needlet space
geometric analysis has a number of advantages that it is immediate
to see (see also \cite{MarVad} for some mathematical results in
this area). For instance, any deviation from the analytic
predictions can be exactly localized on the real and harmonic
space, thus allowing for a much neater interpretation; indeed, a
scale-by-scale probe of asymmetries and relevant features becomes
feasible. Also, while the behaviour of MFs on standard CMB maps is
unavoidably affected by Cosmic Variance, the effect is much smaller 
for MFs evaluated on the highest needlet scales: it becomes
possible to discriminate quite clearly cosmic variance effects
from effective deviations. Indeed, the variances of these
Minkowski functionals converge to zero as the frequency increases,
so that fluctuations around expected values become negligible on
small scales, assuming the null assumptions hold. This allows for
a very precise investigation of asymmetries and anisotropies; in a
future work we shall provide some exact computations on the
variances of these functionals and corresponding aggregated
statistics.

The plan of the paper is as follows: in Section 2, we illustrate
some background material on the Gaussian Kinematic Formula and we
present its application to needlet and multipole components under
the simplest conditions, e.g., full-sky Gaussian maps. In Section
3 we present analytic results for some nonGaussian fields arising
when testing for asymmetries and directional variations in
nonGaussianity, while Section 4 is devoted to the formulae for the
exact expected values in the presence of masked regions. In
Section 5 we present our detailed numerical studies, and we
illustrate our software which allows for numerical corrections of
expected values in the presence of masked regions of any form.
Section 6 draws some conclusions and presents directions for
future work.

\section{The Gaussian Kinematic Formula}

\subsection{The general case}

For cosmological applications, it would seem sufficient to
restrict our attention to random fields or observational data on
the unit sphere $S^{2};$ however we shall show below that
presenting results in a more general setting does yield some
practical advantages, especially when dealing with masked data.
Indeed, the Gaussian Kinematic Formula holds in much greater
generality, and it can certainly be exploited for other
experimental setups, for instance three-dimensional observations
(viewed as data on the three-dimensional ball - this and other
cases will be the object of future works).

On the sphere, the excursion sets $A_{u}(f)$ of a given (possibly random)
function $f$ are defined as
\[
A_{u}(f):=\left\{ x\in S^{2}:f(x)\geq u\right\} \text{ .}
\]%
Of course, in the limit where we take $u=-\infty ,$ we have that $%
A_{u}(f)=S^{2}$.

The \emph{Lipschitz-Killing Curvatures} (LKCs) of these excursion sets,
written
\[
\mathcal{L}_{0}(A_{u}(f)),\mathcal{L}_{1}(A_{u}(f)),\mathcal{L}
_{2}(A_{u}(f))
\]
are defined as:

\begin{itemize}
\item $\mathcal{L}_{0}(A_{u}(f))$ is the Euler-Poincar\'{e}
characteristic, e.g. in two dimensions the number of connected
regions minus the number of holes, and in three dimensions the
number of connected components, minus the number of "handles" plus
the number of holes, see \cite{adlerstflour} for more discussion.
This corresponds to the third Minkowski
functional, or two minus the genus; we recall that the
Euler-Poincar\'{e} characteristic of the full sphere is equal to
two.

\item $\mathcal{L}_{1}(A_{u}(f))$ is half the boundary length of
the excursion regions, e.g. the second Minkowski functional up to
a factor 2. For the full sphere, the boundary length is clearly
zero

\item $\mathcal{L}_{2}(A_{u}(f))$ is the area of the excursion
regions, e.g. the first Minkowski functional. For the full sphere,
one obviously gets $4\pi $.
\end{itemize}

For more general manifolds, the definitions are given in the
Appendix. We shall focus on random fields that have zero mean,
unit variance and are isotropic. These assumptions can be easily
abandoned, entailing just a more complex notation; of course, zero
mean and unit variance can be enforced by normalization
(incidentally, it is well-known that needlet and multipole
components random fields have always zero mean under isotropy).
Let us now introduce some more notation; consider the family of
functions $\rho _{l}(u)$
given by%
\[
\rho _{l}(u)=(2\pi )^{-(l+1)/2}H_{l-1}(u)e^{-u^{2}/2}\text{ ,}
\]%
where $H_{k}(u)$ denotes standard Hermite polynomials, e.g.,
\[
H_{0}(u)=1,
H_{1}(u)=u, H_{2}(u)=u^{2}-1;
\]
we adopt the standard convention that
 \[
H_{-1}(u)=\sqrt{2\pi }(1-\Phi (u))e^{u^{2}/2},
\]
where $\Phi (u)$ is the
standard Gaussian c.d.f., so that%
\begin{eqnarray*}
\rho _{0}(u) &=&(2\pi )^{-1/2}\sqrt{2\pi }(1-\Phi
(u))e^{u^{2}/2}e^{-u^{2}/2}=(1-\Phi (u)) \\
\rho _{1}(u) &=&\frac{1}{2\pi }e^{-u^{2}/2}\text{ , }\rho _{2}(u)=\frac{1}{%
\sqrt{(2\pi )^{3}}}ue^{-u^{2}/2}.
\end{eqnarray*}%
It is interesting to note that $\frac{1}{\sqrt{2\pi }}H_{k}(u)e^{-u^{2}/2}$
gives $(-1)^{k}$ times the $k$-th derivative of a standard Gaussian density,
$k\geq 0.$ In the mathematical literature, this component is written as $%
\mathcal{M}_{l}^{\gamma }([u,\infty ))=\frac{1}{\sqrt{2\pi }}%
H_{k}(u)e^{-u^{2}/2}$ and labelled a Gaussian Minkowski
functional. This terminology, however, may result quite misleading
in a CMB framework, because Gaussian Minkowski functionals are not
at all the same as the Minkowski functionals for Gaussian fields:
hereafter hence we will not use this jargon.

The next ingredient we shall need are the so-called "flag" coefficients,
which are given by%
\[
\left[
\begin{array}{c}
i+l \\
l%
\end{array}%
\right] =\left(
\begin{array}{c}
i+l \\
l%
\end{array}%
\right) \frac{\omega _{i+l}}{\omega _{i}\omega _{l}}\text{ , for }\omega
_{i}=\frac{\pi ^{i/2}}{\Gamma (\frac{i}{2}+1)}\text{ ,}
\]%
so that $\omega _{i}$ represents the area of the $i-$dimensional unit ball, $%
\omega _{1}=2,$ $\omega _{2}=\pi ,$ $\omega _{3}=\frac{4}{3}\pi .$
Finally, we shall introduce a parameter $\lambda$, which
represents the variance of any gradient component at the origin;
equivalently $\lambda$ is simply given by the second derivative of
the covariance function at the origin.

Under these circumstances, for random fields defined on general manifolds $D$
the Gaussian Kinematic Formula is given by the following, extremely elegant
expression (see for instance Theorem 13.2.1 \cite{RFG}:%
\begin{equation}
\lambda ^{i/2}\mathbb{E}\mathcal{L}_{i}(A_{u}(T(x),D))=\sum_{l=0}^{\dim
(D)-i}\left[
\begin{array}{c}
i+l \\
l%
\end{array}%
\right] \lambda ^{(i+l)/2}\rho _{l}(u)\mathcal{L}_{i+l}(D)\text{ .}
\label{GKF}
\end{equation}

This expression may seem unnecessarily complicated, given that in
this paper we shall focus only on spherical random fields: however
this generality will indeed be required below, when we shall
consider masked data (which we will see as data sampled from a
different manifold, i.e. the sphere with sky-cuts).  Before we
proceed, however, it is important to stress some crucial features
of the result given in \eqref{GKF}. Indeed, it must be noted that
the expression on the right-hand side of \eqref{GKF} allows for a
full decoupling of the expected value on the left-hand side into
components which are completely independent: the LKCs of the
original manifold $\mathcal{L}_{k}(D),$ which depend on the
manifold $D$ but not by the threshold value $u$ nor on the
covariance structure of the field we investigate; and the
functions $\rho _{l}(u),$ which depend only on the chosen
threshold level $u$, and are independent from the structure of the
field nor from the properties of the manifold $D$. This will allow
for enormous computational advantages in the sections to follow:
for instance, covering the presence of sky-cuts will entail a new
computation for the values of $\mathcal{L}_{i+l}(D),$ which can be
given once for all for a given mask; this computation will not be
influenced, however, by threshold levels or correlation structure.
Likewise, moving to nonGaussian circumstances will entail a
corresponding replacement of the functions $\rho _{l}(u),$ but no
new computations will be required on correlation structure or to
handle gaps. A particular neat interpretation can be provided, by
simply grouping together the terms $\lambda ^{k/2}\ $\ and
$\mathcal{L}_{k}(D),$ to obtain
\[
\mathcal{L}_{k}^{T}(D)=\lambda ^{k/2}\mathcal{L}_{k}(D)\text{ ;}
\]
in mathematical terms, $\mathcal{L}_{k}^{T}(D)$ is usually described as a
LKC computed with a metric induced by the random field $T,$ e.g. a manifold
which has been rescaled by multiplication times $\sqrt{\lambda },$ the
square root of the second derivative of its covariance function at the
origin. All these notions may seem somewhat abstract, but they yield very
simple analytic expressions in the case of spherical random fields $D=S^{2},$
to which we now turn our attention.

\subsection{The spherical case}\label{ssec:lkc_gauss}

\begin{figure*}
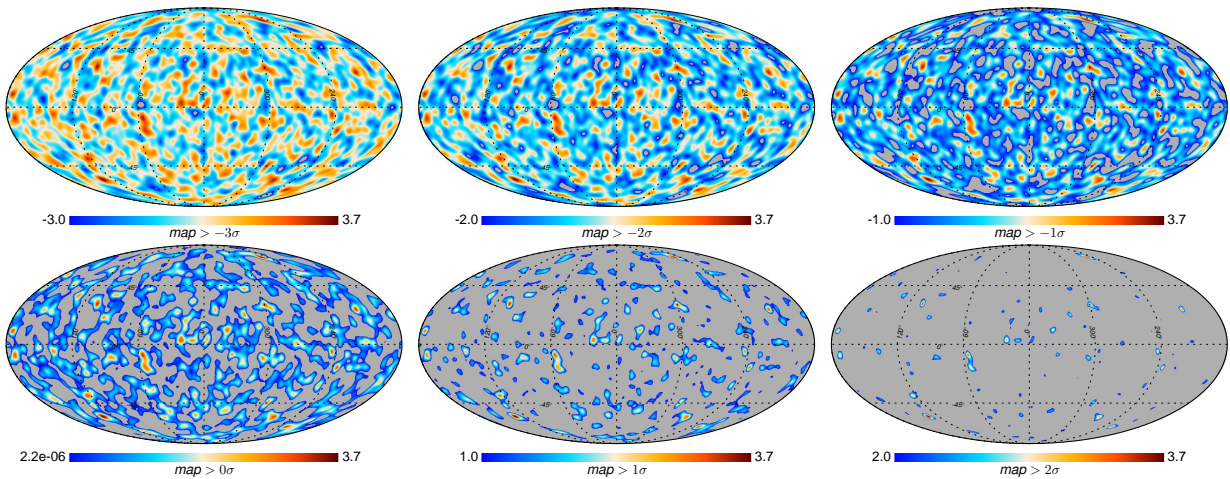
 
\begin{center}
\includegraphics[width=0.3\textwidth,angle=0]{\figdir{map_sup_u-3_fwhm5deg.pdf}}
\includegraphics[width=0.3\textwidth,angle=0]{\figdir{map_sup_u-2_fwhm5deg.pdf}}
\includegraphics[width=0.3\textwidth,angle=0]{\figdir{map_sup_u-1_fwhm5deg.pdf}} \\
\includegraphics[width=0.3\textwidth,angle=0]{\figdir{map_sup_u0_fwhm5deg.pdf}}
\includegraphics[width=0.3\textwidth,angle=0]{\figdir{map_sup_u1_fwhm5deg.pdf}}
\includegraphics[width=0.3\textwidth,angle=0]{\figdir{map_sup_u2_fwhm5deg.pdf}}
\caption{Illustration of excursion fields on a CMB map. The original
  map is smoothed by a $5^\circ$ beam. The subtitles below the color bar
  indicate the threshold levels. \label{fig:eg}}
\end{center}
\end{figure*}

An example of excursion regions of the CMB for different threshold levels is given by \figref{fig:eg}.

The application of the previous general results to the sphere (without
masks) basically provides expression which are already known to the CMB
literature, up to some correction terms. Indeed, for spherical fields $%
\lambda $ it is easily seen to be (see \cite{MarVad})%
\[
\lambda _{\ell }=\frac{\ell (\ell +1)}{2}\text{ , for the multipole field }%
T_{\ell }\text{ , and}
\]%
\[
\frac{\sum_{\ell }b^{4}(\frac{\ell }{2^{j}})C_{\ell }\frac{2\ell +1}{4\pi }%
\frac{\ell (\ell +1)}{2}}{\sum_{\ell }b^{4}(\frac{\ell }{2^{j}})C_{\ell }%
\frac{2\ell +1}{4\pi }}\text{ , for the needlet field }\beta _{j}(x)\text{ ;
}
\]%
note that both fields have been normalized to have unit variance; also, in
this setting%
\[
\left[
\begin{array}{c}
2 \\
0%
\end{array}%
\right] =\left[
\begin{array}{c}
2 \\
2%
\end{array}%
\right] =1\text{ , }\left[
\begin{array}{c}
2 \\
1%
\end{array}%
\right] =\frac{\pi }{2}\text{ .}
\]

Finally, as mentioned earlier the Lipschitz-Killing curvatures
take an extremely simple form on the full sphere: it is indeed
well-known that the Euler-Poincar\'{e} characteristic is
identically equal to 2, the boundary length is of course
zero (the sphere has no boundary), and the area is simply $4\pi ,$ i.e.%
\begin{equation}
\mathcal{L}_{0}(S^{2})=2\text{ , }\mathcal{L}_{1}(S^{2})=0\text{ , }\mathcal{%
L}_{2}(S^{2})=4\pi \text{ .}  \label{spherelkc}
\end{equation}%
After making all these replacements in \eqref{GKF} we thus obtain
general expressions for expected values in the case of multipole
and needlet components which are given in the following two
subsections.

\subsection{Multipole fields}

In the case of a single multipole $T_{\ell }(x)=\sum_{m}a_{\ell m}Y_{\ell
m}(x),$ normalized to have variance one (e.g., divided by $\sqrt{\frac{2\ell
+1}{4\pi }C_{\ell }}),$ the GKF yields immediately%
\begin{equation}
\mathbb{E}\mathcal{L}_{0}(A_{u}(T_{\ell }(.),S^{2}))=2\left\{ 1-\Phi
(u)\right\} +\frac{\ell (\ell +1)}{2}\frac{ue^{-u^{2}/2}}{\sqrt{(2\pi )^{3}}}%
4\pi \text{ ;}  \label{sh1}
\end{equation}
\begin{eqnarray}
\mathbb{E}\mathcal{L}_{1}(A_{u}(T_{\ell }(.),S^{2})) &=&\frac{\pi }{2}%
\left\{ \frac{\ell (\ell +1)}{2}\right\} ^{1/2}\frac{e^{-u^{2}/2}}{2\pi }%
4\pi  \\
&=&\pi \left\{ \frac{\ell (\ell +1)}{2}\right\} ^{1/2}e^{-u^{2}/2}\text{ ;}
\label{sh2}
\end{eqnarray}
and%
\begin{equation}
\mathbb{E}\mathcal{L}_{2}(A_{u}(T_{\ell }(.),S^{2}))=4\pi \times \left\{
1-\Phi (u)\right\} \text{ .}  \label{sh3}
\end{equation}

\subsection{Needlet fields}

The expected value of the Euler-Poincar\'{e} characteristic is given by%
\begin{equation}
\mathbb{E}\mathcal{L}_{0}(A_{u}(\beta _{j}(x),S^{2}))=2\left\{ 1-\Phi
(u)\right\} +\frac{\sum_{\ell }b^{4}(\frac{\ell }{2^{j}})C_{\ell }\frac{%
2\ell +1}{4\pi }P_{\ell }^{\prime }(1)}{\sum_{l}b^{4}(\frac{\ell }{2^{j}}%
)C_{l}\frac{2\ell +1}{4\pi }}\frac{ue^{-u^{2}/2}}{\sqrt{(2\pi )^{3}}}4\pi
\text{ ;}  \label{need11}
\end{equation}
the second Lipschitz-Killing curvature (e.g., half the boundary length) has
expected value%
\begin{equation}
\mathbb{E}\mathcal{L}_{1}(A_{u}(\beta _{j}(x),S^{2}))=\pi \times \left\{
\frac{\sum_{l}b^{4}(\frac{\ell }{2^{j}})C_{l}\frac{2\ell +1}{4\pi }%
P_{l}^{\prime }(1)}{\sum_{l}b^{4}(\frac{\ell }{2^{j}})C_{l}\frac{2\ell +1}{%
4\pi }}\right\} ^{1/2}e^{-u^{2}/2}\text{ ;}  \label{need22}
\end{equation}
Finally, the third Lipschitz-Killing curvature (e.g., the area of the
excursion region) has the following expected value, which is the simplest to
check:%
\begin{equation}
\mathbb{E}\mathcal{L}_{2}(A_{u}(\beta _{j}(x),S^{2}))=4\pi \times \left\{
1-\Phi (u)\right\} \text{ .}  \label{need3}
\end{equation}

The expressions \eqref{sh2}, \eqref{sh3}, \eqref{need22}, \eqref{need3}
match those that would be obtained replacing the angular power
spectrum of a needlet field/multipole component in the standard
expressions for expected values of Minkowski functionals, as given
for instance in \cite{planck2013_IS}, pp.10-11. On the other hand,
on the right-hand side of \eqref{sh1}, \eqref{need11} there is an
extra-term that fully takes into account the spherical geometry:
this term is missing when the result is derived by resorting to a
flat-sky approximation. All these results are perfectly matched by
the simulations presented below; we can hence move to consider
nonGaussian fields and masked regions, as done in the following
Sections.

\begin{figure}
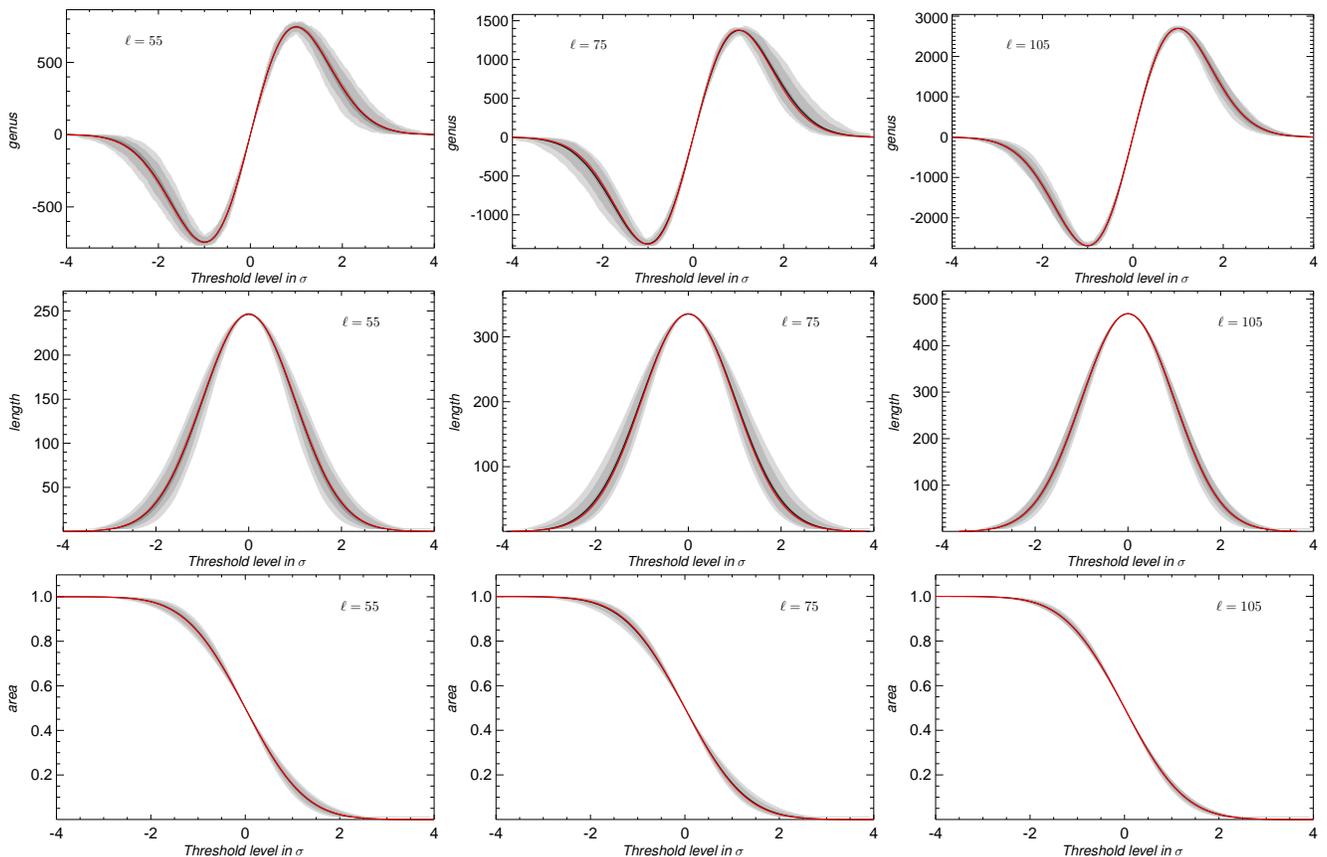
 
\begin{center}
\includegraphics[width=0.32\textwidth,angle=0]{\figdir{genus_mpow1_ell_55_compare.pdf}}
\includegraphics[width=0.32\textwidth,angle=0]{\figdir{genus_mpow1_ell_75_compare.pdf}}
\includegraphics[width=0.32\textwidth,angle=0]{\figdir{genus_mpow1_ell_105_compare.pdf}} \\
\includegraphics[width=0.32\textwidth,angle=0]{\figdir{length_mpow1_ell_55_compare.pdf}}
\includegraphics[width=0.32\textwidth,angle=0]{\figdir{length_mpow1_ell_75_compare.pdf}}
\includegraphics[width=0.32\textwidth,angle=0]{\figdir{length_mpow1_ell_105_compare.pdf}} \\
\includegraphics[width=0.32\textwidth,angle=0]{\figdir{area_mpow1_ell_55_compare.pdf}}
\includegraphics[width=0.32\textwidth,angle=0]{\figdir{area_mpow1_ell_75_compare.pdf}}
\includegraphics[width=0.32\textwidth,angle=0]{\figdir{area_mpow1_ell_105_compare.pdf}} \\
\caption{Multipole space Gaussian case: Analytical (red) vs
  simulations (black and grey). The legend shows the multipoles at
  which the LKCs are evaluated. Grey Shades are $68, 95$ and $99 \%$
  percentiles estimated from 100 simulations. \label{fig:m1ell}}
\end{center}
\end{figure}

\begin{figure}
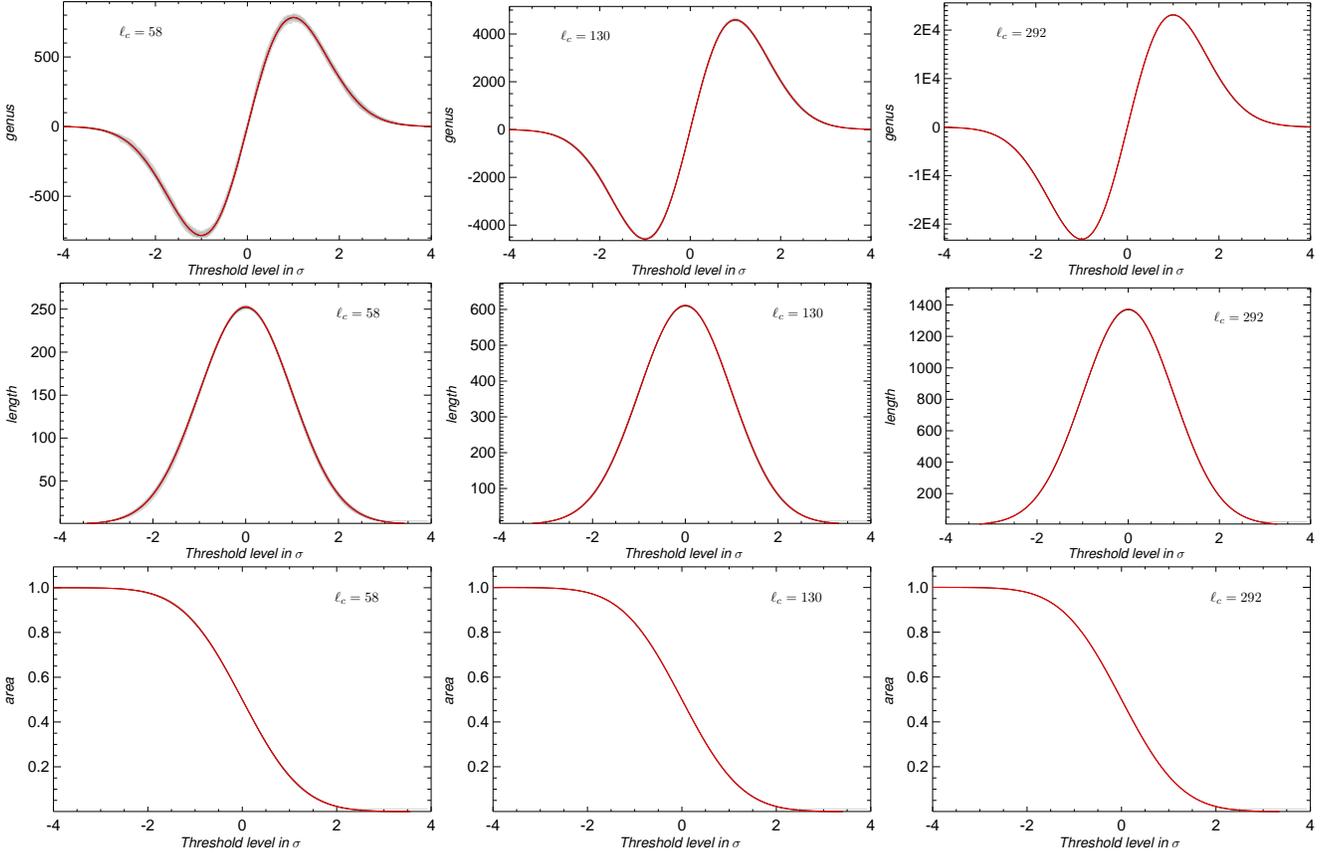
 
\begin{center}
\includegraphics[width=0.32\textwidth,angle=0]{\figdir{genus_mpow1_j_10_compare.pdf}}
\includegraphics[width=0.32\textwidth,angle=0]{\figdir{genus_mpow1_j_12_compare.pdf}}
\includegraphics[width=0.32\textwidth,angle=0]{\figdir{genus_mpow1_j_14_compare.pdf}} \\
\includegraphics[width=0.32\textwidth,angle=0]{\figdir{length_mpow1_j_10_compare.pdf}}
\includegraphics[width=0.32\textwidth,angle=0]{\figdir{length_mpow1_j_12_compare.pdf}}
\includegraphics[width=0.32\textwidth,angle=0]{\figdir{length_mpow1_j_14_compare.pdf}} \\
\includegraphics[width=0.32\textwidth,angle=0]{\figdir{area_mpow1_j_10_compare.pdf}}
\includegraphics[width=0.32\textwidth,angle=0]{\figdir{area_mpow1_j_12_compare.pdf}}
\includegraphics[width=0.32\textwidth,angle=0]{\figdir{area_mpow1_j_14_compare.pdf}} \\
\caption{Needlet space Gaussian case: Analytical (red) vs simulations
  (black and grey). The needlet parameters are $B=1.5$
  $j=10,12,14$. The central multipoles of the corresponding needlet
  filter is given in the legend. Grey Shades are $68, 95$ and $99 \%$
  percentiles estimated from 100 simulations.\label{fig:m1beta}}
\end{center}
\end{figure}

\section{NonGaussian expected values}\label{sec:lkc_ng}

Before we go ahead to discuss the analytic results, it is important to
motivate the class of nonGaussian fields we wish to consider.

A major thread of last decade's research in the field of CMB has been
related to the investigation of possible asymmetries and directional
variations in the observed data; seminal papers in this area were
provided by \cite{Tegmark:2003ve, Vielva:2003et,
  eriksen2004_pa,hansen2004_pa,park2004_mf, Land:2005ad, Larson:2004vm,
  Cruz:2006fy, Copi:2006tu} working on the early WMAP data release,
but the field is still now very active and hotly debated, see
\cite{planck2013_IS} and the references therein. In this
framework, it is well-known that needlet coefficients or fields
can provide unbiased estimates for smoothed versions of the
angular power spectrum, the bispectrum or any higher-order
statistics; these estimates are spatially localized, so they can
be immediately used to test for instance power asymmetries, an
idea first developed in \cite{bkmpBer}, \cite{PBM2}.

More explicitly, consider the squared field $ \beta _{j}^{2}(x);$
from the localization properties of the needlet frame, it is
obvious that the value of $\beta _{j}(x)$ is only determined by
CMB radiation in a small neighbourhood around $x,$ while we have
moreover
\[
\mathbb{E}\beta _{j}^{2}(x)=\mathbb{E}\left\{ \sum_{\ell
=B^{j-1}}^{B^{j+1}}b^{2}(\frac{\ell }{B^{j}})T_{\ell }(x)\right\}
^{2}=\sum_{\ell =B^{j-1}}^{B^{j+1}}b^{4}(\frac{\ell }{B^{j}})\frac{2\ell +1}{%
4\pi }C_{\ell }\text{ ,}
\]%
e.g., the squared coefficients provide natural unbiased estimates for a
binned angular power spectrum. Along the same lines, the cube of these
coefficients provides an unbiased, local estimator of the binned bispectrum,
which is a natural candidate to search for directional variations in
nonGaussianity:%
\begin{eqnarray*}
\mathbb{E}\beta _{j}^{3}(x) &=&\mathbb{E}\left\{ \sum_{\ell
=B^{j-1}}^{B^{j+1}}b^{2}(\frac{\ell }{B^{j}})T_{\ell }(x)\right\} ^{3} \\
&=&\sum_{\ell _{1},\ell _{2},\ell _{3}=B^{j-1}}^{B^{j+1}}b^{2}(\frac{\ell _{1}}{%
B^{j}})b^{2}(\frac{\ell _{2}}{B^{j}})b^{2}(\frac{\ell
_{3}}{B^{j}})\mathbb{E}\left\{ T_{\ell _{1}}(x)T_{\ell
_{2}}(x)T_{\ell _{3}}(x)\right\}
\end{eqnarray*}%
\begin{eqnarray*}
&=&\sum_{\ell _{1},\ell _{2},\ell _{3}=B^{j-1}}^{B^{j+1}}b^{2}(\frac{\ell _{1}}{%
B^{j}})b^{2}(\frac{\ell _{2}}{B^{j}})b^{2}(\frac{\ell
_{3}}{B^{j}})\left(
\begin{array}{ccc}
\ell _{1} & \ell _{2} & \ell _{3} \\
0 & 0 & 0%
\end{array}%
\right) ^{2} \\
&&\times b_{\ell _{1}\ell _{2}\ell _{3}}\sqrt{\frac{(2\ell _{1}+1)(2\ell
_{2}+1)(2\ell _{3}+1)}{4\pi }}\text{ ,}
\end{eqnarray*}%
where $b_{\ell _{1}\ell _{2}\ell _{3}}$ denotes as usual the
reduced bispectrum and the Wigner's 3j symbols have appeared in
the last equation, see \cite{LanM}, \cite{donzelli2012_mfNeedlet}
for more references and details. In the remaining part of this
Section we shall provide the analytic expectation also for the
Minkowski functionals/Lipschitz-Killing curvatures of these cubic
statistics. These results can be rigorously derived by an
application of a more general form of the Gaussian Kinematic
formula, which is given in the Appendix. However, from a more
heuristic point of view their derivation can be provided from a
very simple argument. Indeed, consider for instance a quadratic
transformed field $W=T^{2}:$ the excursion region of the field $W$
over a level $u$ is easily seen to be given by the region where
$T>\sqrt{u},$ plus the region where $T<-\sqrt{u}.$ In view of the
decoupling we reported below, the expected values of the LKCs for
the quadratic case turn out to be just the sum of the
corresponding Gaussian results over these two regions. Likewise,
for the cubic case $W=T^{3}$ the excursion region will be obtained
by simply considering the excursion sets of $T$ over the level
$\sqrt[3]{u}.$ This simple heuristic would not work in more
complicated circumstances where the GKF still provides exact
solutions, but it is enough to justify the results we report
below.

\begin{figure}
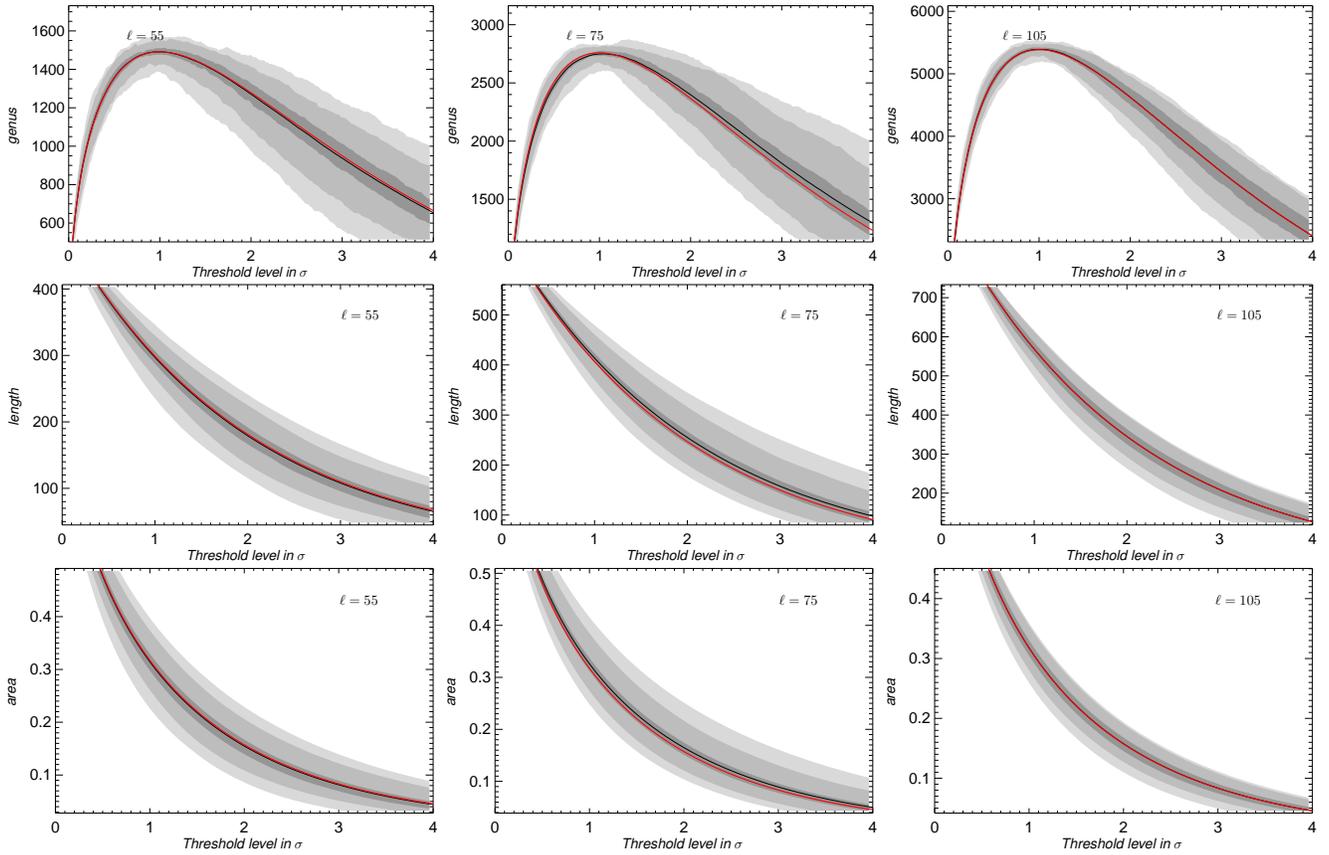
 
\begin{center}
\includegraphics[width=0.32\textwidth,angle=0]{\figdir{genus_mpow2_ell_55_compare.pdf}}
\includegraphics[width=0.32\textwidth,angle=0]{\figdir{genus_mpow2_ell_75_compare.pdf}}
\includegraphics[width=0.32\textwidth,angle=0]{\figdir{genus_mpow2_ell_105_compare.pdf}} \\
\includegraphics[width=0.32\textwidth,angle=0]{\figdir{length_mpow2_ell_55_compare.pdf}}
\includegraphics[width=0.32\textwidth,angle=0]{\figdir{length_mpow2_ell_75_compare.pdf}}
\includegraphics[width=0.32\textwidth,angle=0]{\figdir{length_mpow2_ell_105_compare.pdf}} \\
\includegraphics[width=0.32\textwidth,angle=0]{\figdir{area_mpow2_ell_55_compare.pdf}}
\includegraphics[width=0.32\textwidth,angle=0]{\figdir{area_mpow2_ell_75_compare.pdf}}
\includegraphics[width=0.32\textwidth,angle=0]{\figdir{area_mpow2_ell_105_compare.pdf}} \\
\caption{Multipole space nonGaussian quadratic case: Analytical (red) vs simulations
  (black and grey). The legend shows the multipoles at
  which the LKCs are evaluated. Grey Shades are $68, 95$ and $99 \%$
  percentiles estimated from 100 simulations. \label{fig:m2ell}}
\end{center}
\end{figure}

\begin{figure}
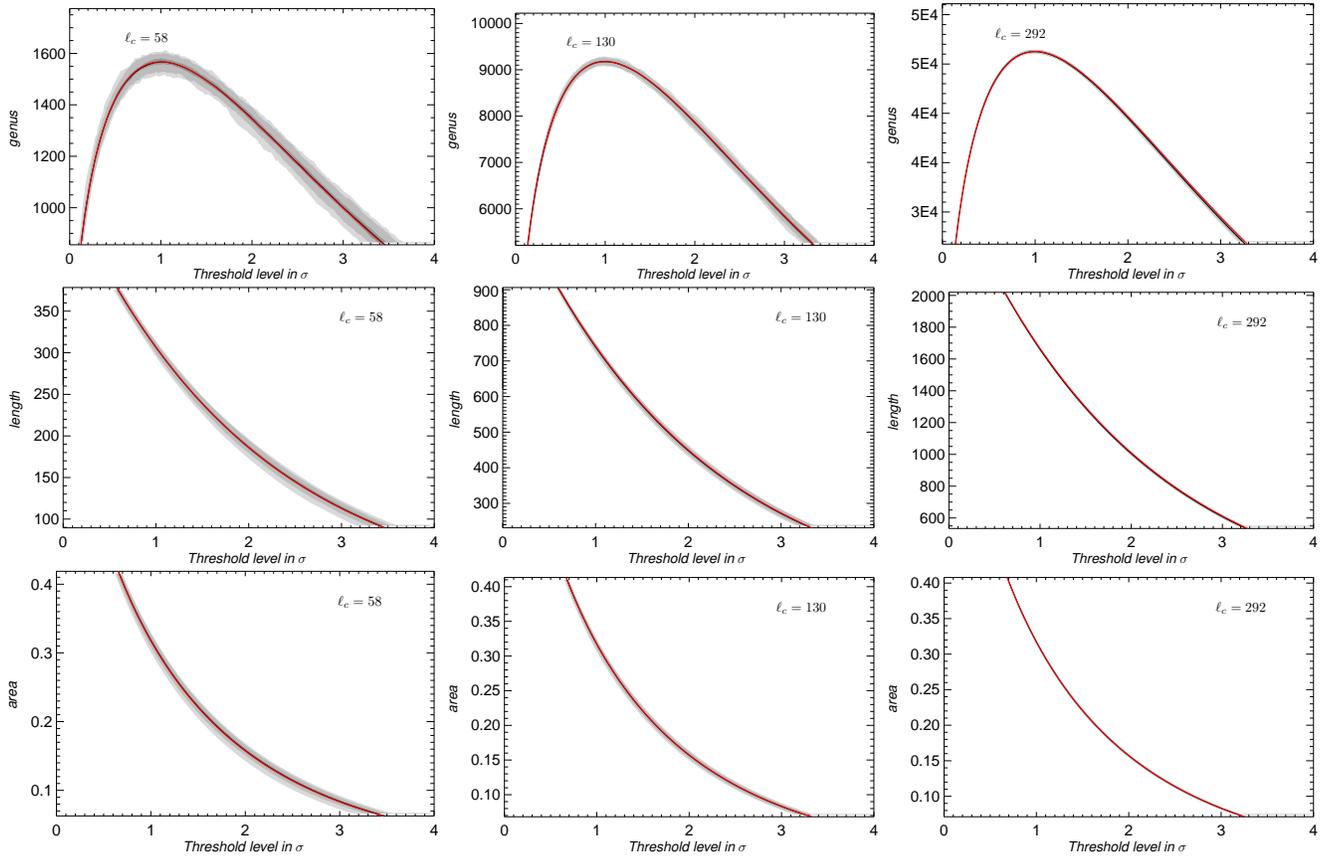
 
\begin{center}
\includegraphics[width=0.32\textwidth,angle=0]{\figdir{genus_mpow2_j_10_compare.pdf}}
\includegraphics[width=0.32\textwidth,angle=0]{\figdir{genus_mpow2_j_12_compare.pdf}}
\includegraphics[width=0.32\textwidth,angle=0]{\figdir{genus_mpow2_j_14_compare.pdf}} \\
\includegraphics[width=0.32\textwidth,angle=0]{\figdir{length_mpow2_j_10_compare.pdf}}
\includegraphics[width=0.32\textwidth,angle=0]{\figdir{length_mpow2_j_12_compare.pdf}}
\includegraphics[width=0.32\textwidth,angle=0]{\figdir{length_mpow2_j_14_compare.pdf}} \\
\includegraphics[width=0.32\textwidth,angle=0]{\figdir{area_mpow2_j_10_compare.pdf}}
\includegraphics[width=0.32\textwidth,angle=0]{\figdir{area_mpow2_j_12_compare.pdf}}
\includegraphics[width=0.32\textwidth,angle=0]{\figdir{area_mpow2_j_14_compare.pdf}} \\
\caption{Needlet space nonGaussian quadratic case: Analytical (red) vs simulations
  (black and grey). The needlet parameters are $B=1.5$
  $j=10,12,14$. The central multipoles of the corresponding needlet
  filter is given in the legend. Grey Shades are $68, 95$ and $99 \%$
  percentiles estimated from 100 simulations.  \label{fig:m2beta}}
\end{center}
\end{figure}

\subsection{The Quadratic case}\label{ssec:lkc_square}

We start from the case where we square the needlet field, as if we were
interested in local estimates of the power spectrum. As usual, we normalize
the starting Gaussian field to have unit variance, and we are hence focusing
on the square field defined by%
\[
\beta _{j,2}(x)=\frac{\beta _{j}^{2}(x)}{Var(\beta _{j}(x))}=\frac{\beta
_{j}^{2}(x)}{\sum_{\ell }b^{4}(\frac{\ell }{B^{j}})\frac{2\ell +1}{4\pi }%
C_{\ell }}\text{ .}
\]

As motivated by the previous heuristic, or as derived more rigorously by the
general Gaussian kinematic formula (see Appendix), we have the following
analytic predictions:

\begin{itemize}
\item For the expected value of the Euler characteristic%
\[
\mathbb{E}\mathcal{L}_{0}(A_{u})=4(1-\Phi
(\sqrt{u}))+4\frac{\sum_{\ell }b^{4}(\frac{\ell
}{B^{j}})\frac{2\ell +1}{4\pi }C_{\ell }\frac{\ell (\ell
+1)}{2}}{\sum_{\ell }b^{4}(\frac{\ell }{B^{j}})\frac{2\ell
+1}{4\pi }C_{\ell }}\frac{e^{-u/2}}{\sqrt{2\pi }}\sqrt{u}\text{ ;}
\]

\item For the second Lipschitz-Killing curvature (i.e., half of the boundary
length)%
\[
\mathbb{E}\mathcal{L}_{1}(A_{u})=2\pi \left\{ \frac{\sum_{\ell }b^{4}(\frac{%
\ell }{B^{j}})\frac{2\ell +1}{4\pi }C_{\ell }\frac{\ell (\ell +1)}{2})}{%
\sum_{\ell }b^{4}(\frac{\ell }{B^{j}})\frac{2\ell +1}{4\pi }C_{\ell }}%
\right\} ^{1/2}e^{-u/2}\text{ ;}
\]

\item Finally for the area of excursion regions%
\[
\mathbb{E}\mathcal{L}_{2}(A_{u})=4\pi \times 2(1-\Phi (\sqrt{u}))\text{ .}
\]
\end{itemize}

The results for the square of normalized multipole components ($T_{\ell
}^{2}/\mathbb{E(}T_{\ell }^{2}))$ are entirely analogous, indeed even
simpler to state:

\begin{itemize}
\item For the expected value of the Euler characteristic%
\[
\mathbb{E}\mathcal{L}_{0}(A_{u})=4(1-\Phi (\sqrt{u}))+4\frac{\ell (\ell +1)}{%
2}\frac{e^{-u/2}}{\sqrt{2\pi }}\sqrt{u}\text{ ;}
\]

\item For the second Lipschitz-Killing curvature (i.e., half of the boundary
length)%
\[
\mathbb{E}\mathcal{L}_{1}(A_{u})=2\pi \left\{ \frac{\ell (\ell +1)}{2}%
\right\} ^{1/2}e^{-u/2}\text{ ;}
\]

\item Finally for the area of excursion regions%
\[
\mathbb{E}\mathcal{L}_{2}(A_{u})=4\pi \times 2(1-\Phi (\sqrt{u}))\text{ .}
\]
\end{itemize}

\begin{figure} 
\begin{center}
\includegraphics[width=0.32\textwidth,angle=0]{\figdir{genus_mpow3_ell_55_compare.pdf}}
\includegraphics[width=0.3\textwidth,angle=0]{\figdir{genus_mpow3_ell_75_compare.pdf}}
\includegraphics[width=0.32\textwidth,angle=0]{\figdir{genus_mpow3_ell_105_compare.pdf}} \\
\includegraphics[width=0.32\textwidth,angle=0]{\figdir{length_mpow3_ell_55_compare.pdf}}
\includegraphics[width=0.32\textwidth,angle=0]{\figdir{length_mpow3_ell_75_compare.pdf}}
\includegraphics[width=0.32\textwidth,angle=0]{\figdir{length_mpow3_ell_105_compare.pdf}} \\
\includegraphics[width=0.32\textwidth,angle=0]{\figdir{area_mpow3_ell_55_compare.pdf}}
\includegraphics[width=0.32\textwidth,angle=0]{\figdir{area_mpow3_ell_75_compare.pdf}}
\includegraphics[width=0.32\textwidth,angle=0]{\figdir{area_mpow3_ell_105_compare.pdf}} \\
\caption{Multipole space nonGaussian cubic case: Analytical (red) vs simulations
  (black and grey). The legend shows the multipoles at
  which the LKCs are evaluated. Grey Shades are $68, 95$ and $99 \%$
  percentiles estimated from 100 simulations.  \label{fig:m3ell}}
\end{center}
\end{figure}

\begin{figure}
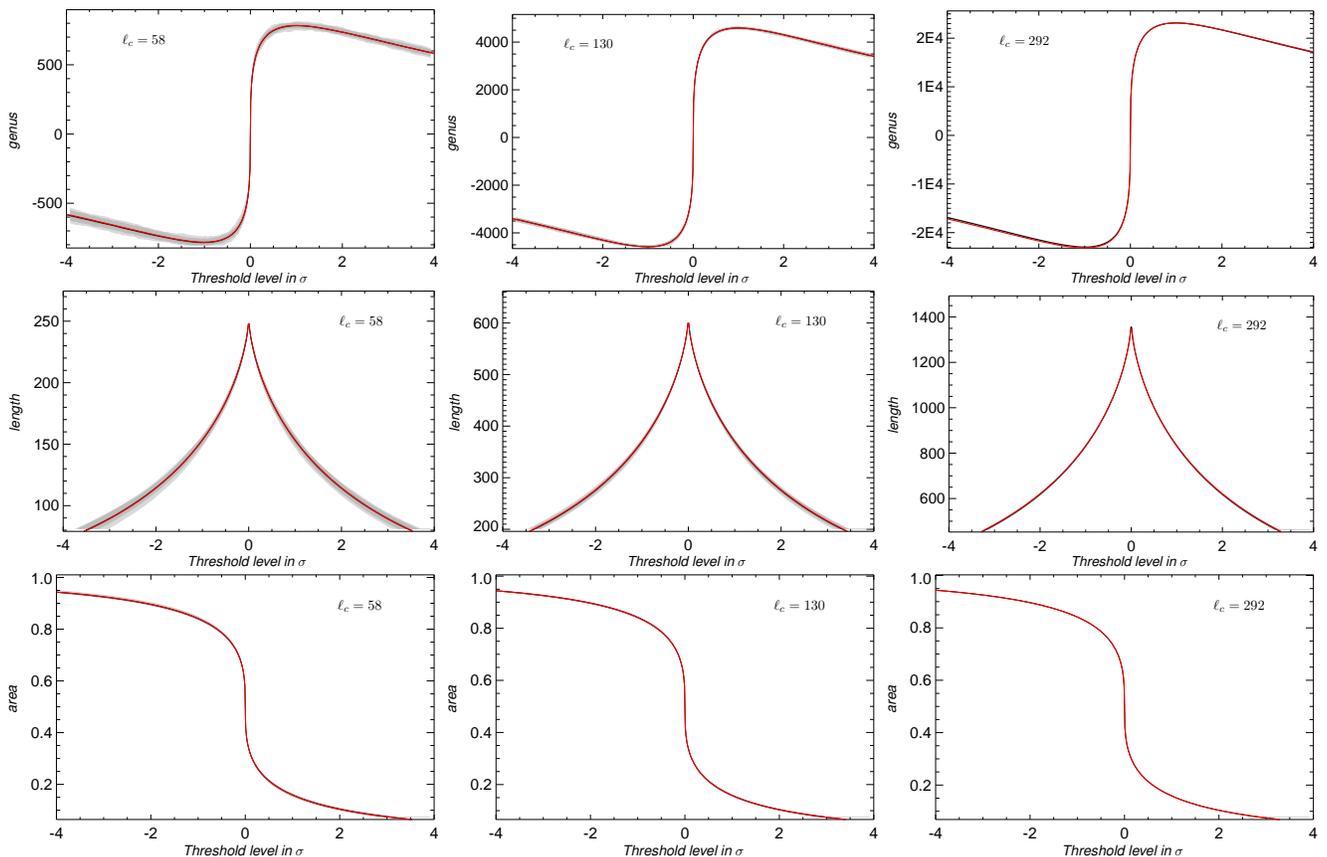
 
\begin{center}
\includegraphics[width=0.32\textwidth,angle=0]{\figdir{genus_mpow3_j_10_compare.pdf}}
\includegraphics[width=0.32\textwidth,angle=0]{\figdir{genus_mpow3_j_12_compare.pdf}}
\includegraphics[width=0.32\textwidth,angle=0]{\figdir{genus_mpow3_j_14_compare.pdf}} \\
\includegraphics[width=0.32\textwidth,angle=0]{\figdir{length_mpow3_j_10_compare.pdf}}
\includegraphics[width=0.32\textwidth,angle=0]{\figdir{length_mpow3_j_12_compare.pdf}}
\includegraphics[width=0.32\textwidth,angle=0]{\figdir{length_mpow3_j_14_compare.pdf}} \\
\includegraphics[width=0.32\textwidth,angle=0]{\figdir{area_mpow3_j_10_compare.pdf}}
\includegraphics[width=0.32\textwidth,angle=0]{\figdir{area_mpow3_j_12_compare.pdf}}
\includegraphics[width=0.32\textwidth,angle=0]{\figdir{area_mpow3_j_14_compare.pdf}} \\
\caption{Needlet space nonGaussian cubic case: Analytical (red) vs simulations
  (black and grey). The needlet parameters are $B=1.5$
  $j=10,12,14$. The central multipoles of the corresponding needlet
  filter is given in the legend. Grey Shades are $68, 95$ and $99 \%$
  percentiles estimated from 100 simulations. \label{fig:m3beta}}
\end{center}
\end{figure}

\subsection{The cubic case $\protect\beta _{j}^{3}(x)$}\label{ssec:lkc_cubic}

Cubic transformations are the natural candidates to search for anisotropies
in the bispectrum, are at least in the skewness; we simply take the cube of
the needlet fields. The analytic prediction are then as follows (see also
\cite{MarVad} and the Appendix for details):

\begin{itemize}
\item The expected value of the Euler characteristic is given by%
\[
\mathbb{E}\mathcal{L}_{0}(A_{u}(\beta _{j}^{3}(x);S^{2}))=2(1-\Phi (\sqrt[3]{%
u}))+2\frac{\sum_{\ell }b^{4}(\frac{\ell }{B^{j}})\frac{2\ell +1}{4\pi }%
C_{\ell }\frac{\ell (\ell +1)}{2}}{\sum_{\ell }b^{4}(\frac{\ell }{B^{j}})%
\frac{2\ell +1}{4\pi }C_{\ell }}\frac{e^{-(\sqrt[3]{u})^{2}/2}}{\sqrt{2\pi }}%
\sqrt[3]{u}\text{ ;}
\]

\item The expected value for half the boundary length is%
\[
\mathbb{E}\mathcal{L}_{1}(A_{u}(\beta _{j}^{3}(x);S^{2}))=\pi \left\{ \frac{%
\sum_{l}b^{4}(\frac{\ell }{B^{j}})\frac{2\ell +1}{4\pi }C_{\ell
}\frac{\ell
(\ell +1)}{2}}{\sum_{\ell }b^{4}(\frac{\ell }{B^{j}})\frac{2\ell +1}{4\pi }%
C_{\ell }}\right\} ^{1/2}e^{-(\sqrt[3]{u})^{2}/2}\text{ ;}
\]

\item Finally, the expected value of the area of excursion regions is
\[
\mathbb{E}\mathcal{L}_{2}(A_{u}(\beta _{j}^{3}(x);S^{2}))=4\pi (1-\Phi (\sqrt%
[3]{u}))\text{ .}
\]
\end{itemize}

The corresponding values for the cube of normalized multipole components are
given by

\begin{itemize}
\item The expected value of the Euler characteristic is given by%
\[
\mathbb{E}\mathcal{L}_{0}(A_{u}(\beta _{j}^{3}(x);S^{2}))=2(1-\Phi (\sqrt[3]{%
u}))+2\frac{\ell (\ell +1)}{2}\frac{e^{-(\sqrt[3]{u})^{2}/2}}{\sqrt{2\pi }}%
\sqrt[3]{u}\text{ ;}
\]

\item The expected value for half the boundary length is%
\[
\mathbb{E}\mathcal{L}_{1}(A_{u}(\beta _{j}^{3}(x);S^{2}))=\pi \left\{ \frac{%
\ell (\ell +1)}{2}\right\} ^{1/2}e^{-(\sqrt[3]{u})^{2}/2}\text{ ;}
\]

\item Finally, the expected value of the area of excursion regions is
\[
\mathbb{E}\mathcal{L}_{2}(A_{u}(\beta _{j}^{3}(x);S^{2}))=4\pi (1-\Phi (\sqrt%
[3]{u}))\text{ .}
\]
\end{itemize}

It should be noted that the area measure is completely insensitive to the
behaviour of the correlation structure, and therefore takes the same values
in the needlet and multipole cases.

We recall that in \cite{MarVad} further nonGaussian cases have been
considered, e.g. the situation where the polynomial transforms of these
coefficients are further averaged by moving disks centred at varying pixels
on the sphere. Analytical results have been provided even for these
circumstances, however for brevity's sake we delay their investigation to
future research.

\begin{figure} 
\begin{center}
\includegraphics[width=0.6\textwidth,angle=0]{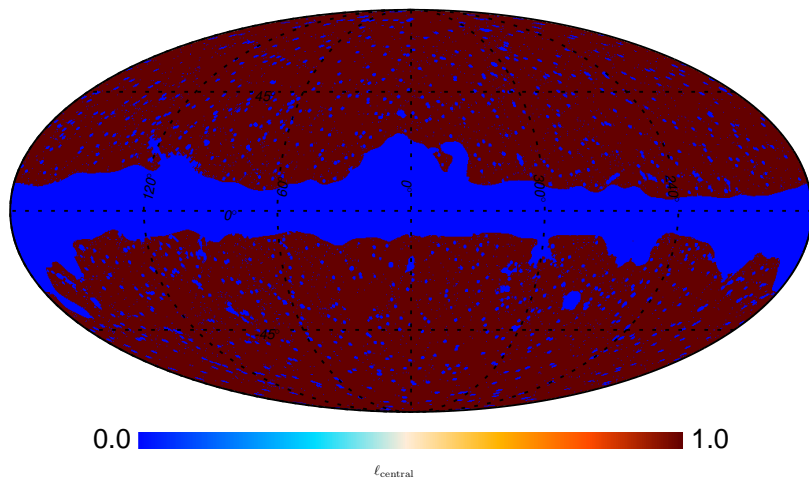}
\caption{Planck Union mask. The fractional area of the unmasked region is $f_{sky}=0.73$. \label{fig:u73mask}}
\end{center}
\end{figure}

\section{Masked Regions}\label{sec:lkc_masked}

In the analysis of data collected from experiments with masked
regions, as it is basically always the case in Cosmology, the full
power of the GKF emerges most clearly. Let us denote by
$M=S^{2}\backslash G$ the sphere to which the masked regions (for
instance, the galactic cut) have been subtracted; it is
then sufficient to replace the LKCs $\mathcal{L}_{i+l}(M)$ to $\mathcal{L}%
_{i+l}(S^{2})$ in \eqref{GKF},\eqref{spherelkc} to obtain the
desired result. At first sight, however, this may appear as a very
difficult task; how to replace the simple values provided in
\eqref{spherelkc} with the LKC for a masked region, possibly with
a highly complicated structure including many removed point
sources and other foreground regions with complex shapes? For the
area measure $\mathcal{L}_{2}(S^{2}\backslash G)$ the computation
could be trivial (by simply adjusting the sky fraction), but for
the boundary length $\mathcal{L}_{1}(S^{2}\backslash G)$ and the
Euler-Poincar\'{e} characteristic $\mathcal{L}_{0}(S^{2}\backslash
G)$ this problem may seem quite hard, especially when a huge
number of removed point sources is given.

A very simple solution can however be provided by exploiting one
more time Gaussian Kinematic Formula, following an idea discussed in
\cite{adlerstflour}, chapter 5.4. In fact, for any given mask one can choose a
simple isotropic random field with known angular power spectrum,
and from this one may evaluate by Monte Carlo simulations the
realized values of LKC of excursion sets at some fixed levels of
threshold values $u$. These realized values can then be compared
with the analytic predictions; for a given input angular power
spectrum, these are fully known, up to some fixed parameters
representing the LKCs $ \mathcal{L}_{i}(S^{2}\backslash G)$. These
parameters can then be estimated once for all by simple least
square regression, and used as an input to derive analytic
predictions for a given mask. These predictions would hold for
arbitrary threshold values $u$ and irrespective of the covariance
structure, the frequency or scales $j,\ell $ considered, the
Gaussian or nonGaussian circumstances.

In summary, the following multi-step procedure is advocated:

\begin{enumerate}
\item Fix a simple power spectrum $C_{\ell },$ for instance with $L_{\max }=10,$
and generate Gaussian maps out of it 

\item Fix a limited number of threshold values $u$ and perform a
Monte Carlo evaluation of the LKCs evaluated on the excursion set
of the fields generated according to 1

\item Use least square regression to estimate
$\mathcal{L}_{i}(S^{2}\backslash G)$, $i=0,1,2$ in equation
\eqref{GKF}

\item Use the estimates obtained in point 3 as an input for equation
\eqref{GKF} for any arbitrary power spectrum (for instance,
multipole or needlet components on realizations of a $\Lambda CDM$
model, under Gaussian and nonGaussian circumstances).
\end{enumerate}

We believe that this routine illustrates very vividly the
advantages of the decoupling between domain manifold, covariance
structure and threshold value achieved by the Gaussian Kinematic
Formula \eqref{GKF}. The resulting predictions are indeed
extremely accurate, as illustrated in the following Section.

\begin{figure}
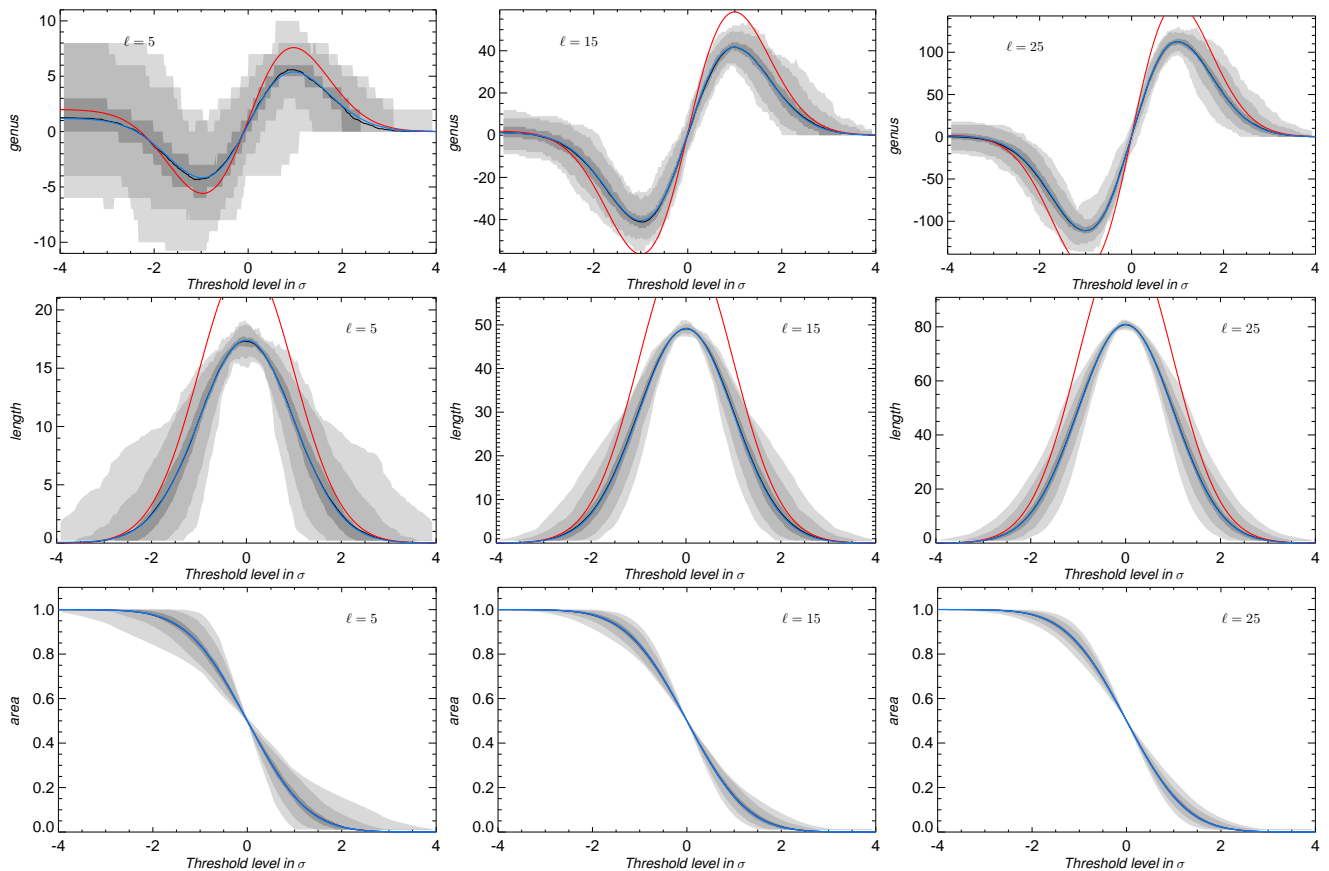
 
\begin{center}
\includegraphics[width=0.32\textwidth,angle=0]{\figdir{genus_mpow1_ell_5_compare_planck.pdf}}
\includegraphics[width=0.32\textwidth,angle=0]{\figdir{genus_mpow1_ell_15_compare_planck.pdf}}
\includegraphics[width=0.32\textwidth,angle=0]{\figdir{genus_mpow1_ell_25_compare_planck.pdf}} \\
\includegraphics[width=0.32\textwidth,angle=0]{\figdir{length_mpow1_ell_5_compare_planck.pdf}}
\includegraphics[width=0.32\textwidth,angle=0]{\figdir{length_mpow1_ell_15_compare_planck.pdf}}
\includegraphics[width=0.32\textwidth,angle=0]{\figdir{length_mpow1_ell_25_compare_planck.pdf}} \\
\includegraphics[width=0.32\textwidth,angle=0]{\figdir{area_mpow1_ell_5_compare_planck.pdf}}
\includegraphics[width=0.32\textwidth,angle=0]{\figdir{area_mpow1_ell_15_compare_planck.pdf}}
\includegraphics[width=0.32\textwidth,angle=0]{\figdir{area_mpow1_ell_25_compare_planck.pdf}} \\
\caption{Multipole space Gaussian masked case: Analytical (red - full sky;  blue - mask corrected) vs simulations
  (black and grey). The legend shows the multipoles at
  which the LKCs are evaluated. Grey Shades are $68, 95$ and $99 \%$
  percentiles estimated from 100 simulations. \label{fig:m1ell_mask}}
\end{center}
\end{figure}

\begin{figure}
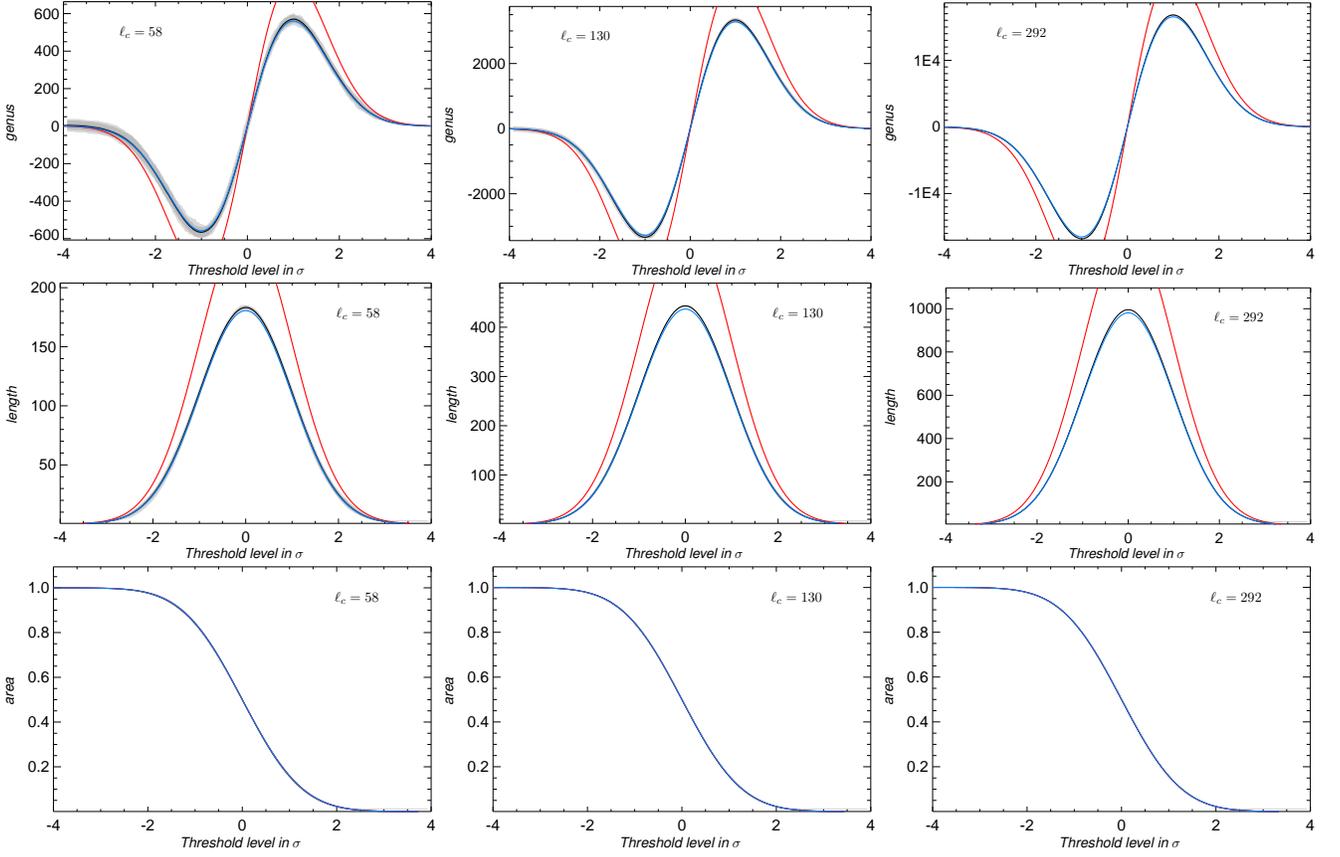
 
\begin{center}
\includegraphics[width=0.32\textwidth,angle=0]{\figdir{genus_mpow1_j_10_compare_planck.pdf}}
\includegraphics[width=0.32\textwidth,angle=0]{\figdir{genus_mpow1_j_12_compare_planck.pdf}}
\includegraphics[width=0.32\textwidth,angle=0]{\figdir{genus_mpow1_j_14_compare_planck.pdf}} \\
\includegraphics[width=0.32\textwidth,angle=0]{\figdir{length_mpow1_j_10_compare_planck.pdf}}
\includegraphics[width=0.32\textwidth,angle=0]{\figdir{length_mpow1_j_12_compare_planck.pdf}}
\includegraphics[width=0.32\textwidth,angle=0]{\figdir{length_mpow1_j_14_compare_planck.pdf}} \\
\includegraphics[width=0.32\textwidth,angle=0]{\figdir{area_mpow1_j_10_compare_planck.pdf}}
\includegraphics[width=0.32\textwidth,angle=0]{\figdir{area_mpow1_j_12_compare_planck.pdf}}
\includegraphics[width=0.32\textwidth,angle=0]{\figdir{area_mpow1_j_14_compare_planck.pdf}} \\
\caption{Needlet space nonGaussian masked case: Analytical (red - full sky;  blue - mask corrected) vs simulations
  (black and grey). The needlet parameters are $B=1.5$
  $j=10,12,14$. The central multipoles of the corresponding needlet
  filter is given in the legend. Grey Shades are $68, 95$ and $99 \%$
  percentiles estimated from 100 simulations. \label{fig:m1beta_mask}}
\end{center}
\end{figure}

\section{Numerical results}\label{sec:numerical}
In this section we describe the comparison of the analytical
results outlined in the previous sections to the corresponding
results from simulations.  In all cases we generated 100 map
realizations of an input power spectrum using the
\healpix\cite{healpix} package. We estimated LKCs from each
simulations and compared their mean with the analytical results.
We found an excellent agreement in all the cases that we
investigated; more precisely, not only the estimated curves are
always well within the $68\%$ Confidence Interval (CL), but
actually as shown below they are for practical purposes basically
indistinguishable from the theoretical predictions even with a
relatively low number of Monte Carlo simulations.

\subsection*{Simulations and Algorithm}
We used \healpix \emph{synfast} to simulate a map from a given
power spectrum; the choice of this power spectrum has no influence
on the results we shall provide.  The procedures to obtain the
single multipole or needlet maps are standard and can be described
as follows: first we harmonic transform the simulated maps using
\emph{anafast}; then to obtain $T_\ell(x)$ or $\beta_j(x)$ maps,
we simply take the appropriate inverse transform across the
relevant multipoles, in the case of needlets inserting also the
squared needlet filter $b^{2}(.)$. The multipole/needlet maps are
then normalized by their root mean square, which is computed
analytically using the input power spectrum, see below.

From these normalized multipole/needlet maps we then computed the
three Minkowski Functionals, which as argued earlier are
equivalent to the LKCs up to constant factors.  This
implementation is achieved exploiting the algorithms described in
\cite{eriksen2004_mf}. In short, these algorithms can be described
as follows: the area, i.e. the first MF, is computed by evaluating
the number of pixels above a certain threshold. The length, the
second MF, is computed by tracing isocontour lines in pixel space.
For a sufficiently high-resolution map, pixels around isocontour
lines have different signs relative to the contour line, after
normalizing the lines to zero. To measure the length of these
lines, sets of four pixels are compared; when at least two of them
have different signs, the locations where the contour line enters
and exits these sets of pixels are determined and the length is
iteratively calculated by standard dot product. The
Euler-Poincare\'{e}, the third MF, is computed by means of its
characterizations through Morse theory; more explicitly, critical
points are determined as the pixels where the gradient vanishes.
The Hessian matrices around these critical points are computed,
and their so-called indexes (i.e., the sign of their determinant,
or the product of their eigenvalues) are evaluated. Positive
indexes correspond to extrema (minima plus maxima), negative
indexes to saddles; in two dimensions, the Euler-Poincar\'{e}
characteristic is simply obtained as the difference between the
number of extrema and the number of saddles.

\subsection*{On normalization issues}
As mentioned, all the maps we used to estimate the LKCs are
normalized to have unit variance; hence the threshold levels
$-2,-1,0,1,2,\dots$ are given in terms of the standard deviation.
It should be noted that at low multipoles, the sample variance
need not be close to the population value, due to Cosmic Variance
effect.  As a result of this, normalizing maps by their respective
sample \rms would lead to incorrect estimates of the mean and
variance of LKCs. We also stress that population variances can
trivially be derived from any given power spectrum; for instance,
the variance of a needlet map at frequency $j$ is given by
\[
\sigma^2(\beta_j) = \sum_{\ell=B^{j-1}}^{B^{j-1}}{b^4(\frac{\ell}{B^j})\frac{(2\ell+1)C_l}{4\pi}}
\]
In the case where the input spectra are not known, one should use the
best-fit power spectra from the map to compute the normalization factor.

\subsection*{Code validation}
To understand the accuracy of our code in estimating the MFs,
in particular in measuring the length of isocontour lines,
we used some test functions for which the relevant quantities are analytically known.
For instance one such function we used is
\[
f(\theta,\phi) = sin(n\theta),
\]
for which the length of isocontour lines at level zero are given by
$2\pi\sum_{k=1}^{n-1}sin(\frac{k}{n}\pi)$;
the results from our code are consistent with these theoretical values
to better than $0.001\%$. Of course, the accuracy may degrade for highly
oscillatory functions, but we believe this test provides a good
validation to the entire pipeline and shows that the algorithms we
employed are very reliable.

\subsection*{Results: Gaussian fields}

In \figref{fig:m1ell} we compare the multipole space analytical
results (red curve) given in \secref{ssec:lkc_gauss} with that of
the simulations (black curve - mean of the simulations). The
$68\%,95\%$ and $99\%$ CLs are shown from dark to light grey
bounds. From left to right panels, the plots shows the results
corresponding to multipoles $\ell=5,50,105$. We stress that our
fit is extremely accurate, even at very low multipole values where
the flat-sky approximation which is usually adopted cannot be
expected to hold.  We also note the improved concentration around
the expected values at higher-multipoles; indeed, the same
behaviour of these variances can be predicted analytically, but we
delay these results for future work.

Likewise, \figref{fig:m1beta} shows analogous results in needlet
space; the colors for different curves have the same meaning as
described above. The displayed results cover the frequencies
$j=10,12,14$  which for $B=1.5$  correspond to multipoles in the order of
60,130,200; these results are even more accurate than in the
multipole case, in particular the decay of Cosmic Variance is
faster.


\subsection*{Results: Non-Gaussian fields}
As described before, our non-Gaussian maps are constructed by
taking a power transform of a Gaussian input. We also argued
earlier in \secref{sec:lkc_ng} that the quadratic power transform
seems useful to investigate power spectrum asymmetries, while the
cubic transform provides a natural probe of possible directional
variations in non-Gaussianity.

In \figref{fig:m2ell} and \figref{fig:m2beta} we compare the analytical results
for the quadratic case in multipole and needlet space with those
from the simulations, respectively. As described above the red
curves are for analytical predictions, while the black and grey
ones are for simulations.

Similarly, in \figref{fig:m3ell} and \figref{fig:m3beta} we show analytical
vs. simulation results for the cubic nonGaussian case in multipole and
needlet space, respectively. The fit between predicted values and
simulations is again extremely good.

\subsection*{Results: masked sky case}
Probably the main contribution in this paper relates to the
possibility to use the GKF to handle analytically the effect of
sky cuts on Minkowski functionals, see the discussion in
\secref{sec:lkc_masked}.  As a numerical validation of the
analytical results for the expected values of LKCs in the presence
of sky-mask, here we use the realistic Planck official sky mask,
which is formed as a union of different foreground separation
methods confidence masks together with point source masks. The cut
regions are shown in \figref{fig:u73mask}, leaving on observed
area of $f_{\rm sky} = 73\%$. As explained earlier, the key step
is the evaluation of LKCs for the masked sphere, which can then be
used as input values to predict the LKCs of excursion sets under
arbitrary covariance structures. In particular, the input LKCs for
the masked sphere have been derived by simulation from a masked
single multipole field at $\ell=15$, map; this ensures that the
estimation procedure can be implemented with remarkable
computational efficiency. The resulting values are then inserted
to obtain the analytic predictions at any frequency or multipole.

In \figref{fig:m1ell_mask} and \figref{fig:m1beta_mask} we compare the masked
Gaussian field analytical result with the corresponding
simulations in multipole and needlet space, respectively. Of
course, here as for the other cases the most relevant results in
practice are those for needlets, because single multipoles cannot
be extracted from masked data; nevertheless, it is reassuring that
the fit works in both circumstances. Moreover, the analysis of
multipole components can be exploited to verify the statistical
properties of full-sky maps, as those obtained for instance by
means of inpainting techniques. These issues are left as topics
for further research.

\section{Summary and Conclusion}

In this paper, we illustrated a number of applications for
Cosmological data analysis of the Gaussian Kinematic Formula
(GKF), (see \cite {TaylorAdler2003, Taylor2006, TaylorAdler2009},
\cite{adlerstflour}, \cite{RFG}). The Gaussian Kinematic Formula
allows to evaluate exact expected values for Lipschitz-Killing
curvatures (Minkowski functionals) in a number of circumstances of
applied interest, covering in particular full-sky experiments
(accounting for the geometry of the sphere), nonlinear statistics
and masked data.

We used the GKF on random fields derived
by harmonic and needlet transforms, allowing for the further advantage
of better control of Cosmic Variance effects and localization.  In
particular we provided the analytic expressions for the Minkowski
functionals for needlets and single multipole fields, covering
Gaussian and nonGaussian circumstances, with and without masks.  All
the results reported are validated by an extensive Monte Carlo study,
which demonstrates an extremely good agreement between predictions and
simulations.

\section{Acknowledgments}
The authors acknowledge support from ERC Grant 277742 Pascal. 
We acknowledge the use of resources from the
Norwegian national super-computing facilities, NOTUR. Maps and results
have been derived using the \healpix (http://healpix.jpl.nasa.gov)
software package developed by \cite{healpix}.


\bibliography{mf_literature_wabstract}

\section{Mathematical Appendix}

On a general, high-dimensional manifold, the LKCs for the region $A$ are
defined as the coefficients of a Taylor expansion of a \emph{Tube} of radius
$r$ around $A.$ Formally, a Tube is simply the set $A$ plus an halo, i.e.%
\[
Tube(A,r)=\left\{ x:d(x,A)\leq r\right\} \text{ .}
\]%
assuming that $A$ had dimension $\dim (A)=n,$ the LKCs are implicitly
defined by the formula%
\[
Vol\left[ Tube(A,r)\right] =\sum_{k=0}^{n}\mathcal{L}_{n-k}(A)\omega
_{k}r^{k}.
\]%
For instance, let $A$ be the unit square on the plane; by elementary
geometry, the volume of the Tube is then given by%
\[
\mathcal{L}_{2}(A)+2\mathcal{L}_{1}(A)r+\mathcal{L}_{0}(A)\pi r^{2}=1+2\cdot
2\cdot r+\pi r^{2},
\]%
whence it is seen that in the two-dimensional case the LKCs correspond to
Euler-Poincar\'{e} characteristic, half the boundary length and area,
respectively. This definition extends to arbitrary manifolds and dimensions,
and makes it possible to express the GKF in much greater generality.
Similarly, one can introduce the Gaussian Minkowski functionals $\mathcal{M}%
_{k}(U)$ as the Taylor coefficients in the expansion of the Tube
probabilities, e.g%
\[
\Pr \left\{ Z\in Tube(U,r)\right\} =\sum_{k}\mathcal{M}_{k}(U)\frac{r^{k}}{k!}.
\]%
The left-hand side simply represents the probability that a zero-mean
standard Gaussian variable belongs to $Tube(U,r);$ for instance, for $%
U=[u,\infty )$ it can be checked that the Gaussian Minkowski functionals
yield the $k-$order derivatives of Gaussian densities that we recalled
above. More general forms of $U$ are necessary, however, when one considers
nonGaussian processes, as we shall do below.

We shall now discuss the Gaussian kinematic formula for the case of
nonlinear transforms of Gaussian and isotropic random fields; i.e., we shall
consider fields of the form%
\[
y(x)=g(T(x))\text{ , }
\]%
where $T(x)$ is zero-mean, unit variance, Gaussian and isotropic, and the
function $g(.)$ is such that also $y(.)$ has finite variance; for our
purposes, the examples we shall consider are simply quadratic and cubic
polynomials, i.e. $g(T)=T^{2}$ and $g(T)=T^{3}.$ Under these circumstances,
the Gaussian kinematic formula takes the form%
\begin{equation}
\lambda ^{i/2}\mathbb{E}\mathcal{L}_{i}(A_{u}(g(T),M))=\sum_{k=0}^{\dim
(M)-i}\lambda ^{(i+k)/2}\mathcal{L}_{i+k}(M)\mathcal{M}_{k}(g^{-1}[u,\infty
))\text{ ;}  \label{GKF2}
\end{equation}%
the expression obviously becomes identical to \eqref{GKF}, in the
Gaussian case $g(T(x))=T(x).$ For more general transforms, the
role of the Gaussian Minkowski functionals becomes crucial: these
are rather simple to evaluate for quadratic and cubic cases, as we
shall show below.

\subsection{The Quadratic Case}

Here we are interested in the analysis of quadratic functionals such as
\[
g(\beta _{j}(x))=\frac{\beta _{j}^{2}(x)}{\mathbb{E}\beta _{j}^{2}(x)}\text{
.}
\]%
By the general Gaussian kinematic formula and simple computations we have%
\begin{eqnarray*}
\mathbb{E}\mathcal{L}_{0}(A_{u}(g(\beta _{j}(x)),S^{2})
&=&\sum_{k=0}^{2}(2\pi )^{-k/2}\lambda _{j}^{k/2}\mathcal{L}_{k}(S^{2})%
\mathcal{M}_{k}((-\infty ,-\sqrt{u})\cup (\sqrt{u},\infty )) \\
&=&\sum_{k=0}^{2}(2\pi )^{-k/2}\lambda _{j}^{k/2}\mathcal{L}_{k}(S^{2})2%
\mathcal{M}_{k}^{\mathcal{N}}((\sqrt{u},\infty ))
\end{eqnarray*}%
\[
=2\cdot 2\cdot (1-\Phi (\sqrt{u}))+0+\frac{1}{2\pi }\mathcal{L}_{2}^{\beta
_{j}}(S^{2})\frac{e^{-u/2}}{\sqrt{2\pi }}2\sqrt{u}
\]%
\[
=2\cdot 2\cdot (1-\Phi (\sqrt{u}))+\frac{1}{2\pi }\frac{\sum_{\ell }b^{2}(%
\frac{\ell }{B^{j}})\frac{2\ell +1}{4\pi }C_{\ell }\frac{\ell (\ell +1)}{2}}{%
\sum_{\ell }b^{2}(\frac{\ell }{B^{j}})\frac{2\ell +1}{4\pi }C_{\ell }}%
\mathcal{L}_{2}(S^{2})\frac{e^{-u/2}}{\sqrt{2\pi }}2\sqrt{u}\text{ .}
\]%
Also%
\begin{eqnarray*}
\lambda _{j}^{1/2}\mathbb{E}\mathcal{L}_{1}((A_{u}(g(\beta _{j}(x)),S^{2}))
&=&\sum_{k=0}^{1}(2\pi )^{-k/2}\left[
\begin{array}{c}
k+1 \\
k%
\end{array}%
\right] \lambda _{j}^{(k+1)/2}\mathcal{L}_{k+1}(S^{2})\mathcal{M}%
_{k}(g^{-1}[u,\infty )) \\
&=&\lambda _{j}^{1/2}\mathcal{L}_{1}(S^{2})\mathcal{M}_{0}((-\infty ,-\sqrt{u%
})\cup (\sqrt{u},\infty )) \\
&&+(2\pi )^{-1/2}\frac{\pi }{2}\lambda _{j}\mathcal{L}_{2}(S^{2})\mathcal{M}%
_{1}^{\mathcal{N}}((-\infty ,-\sqrt{u})\cup (\sqrt{u},\infty )) \\
&=&(2\pi )^{-1/2}\frac{\pi }{2}(4\pi \frac{\sum_{\ell }b^{2}(\frac{\ell }{%
B^{j}})\frac{2\ell +1}{4\pi }C_{\ell }\frac{\ell (\ell +1)}{2}}{\sum_{\ell
}b^{2}(\frac{\ell }{B^{j}})\frac{2\ell +1}{4\pi }C_{\ell }}\mathcal{)}2\frac{%
e^{-u/2}}{\sqrt{2\pi }} \\
&=&2\pi (\frac{\sum_{\ell }b^{2}(\frac{\ell }{B^{j}})\frac{2\ell +1}{4\pi }%
C_{\ell }\frac{\ell (\ell +1)}{2}}{\sum_{\ell }b^{2}(\frac{\ell }{B^{j}})%
\frac{2\ell +1}{4\pi }C_{\ell }}\mathcal{)}e^{-u/2},
\end{eqnarray*}%
which implies%
\[
\mathbb{E}\mathcal{L}_{1}((A_{u}(g(\beta _{j}(x)),S^{2}))=2\pi \left\{ \frac{%
\sum_{\ell }b^{2}(\frac{\ell }{B^{j}})\frac{2\ell +1}{4\pi }C_{\ell }\frac{%
\ell (\ell +1)}{2}}{\sum_{\ell }b^{2}(\frac{\ell }{B^{j}})\frac{2\ell +1}{%
4\pi }C_{\ell }}\right\} ^{1/2}e^{-u/2}
\]%
entailing a length of the boundary of excursion sets given by%
\[
2\pi \left\{ \frac{\sum_{\ell }b^{2}(\frac{\ell }{B^{j}})\frac{2\ell +1}{%
4\pi }C_{\ell }\frac{\ell (\ell +1)}{2}}{\sum_{\ell }b^{2}(\frac{\ell }{B^{j}%
})\frac{2\ell +1}{4\pi }C_{\ell }}\right\} ^{1/2}e^{-u/2}.
\]%
Finally%
\begin{eqnarray*}
\lambda _{j}\mathbb{E}\mathcal{L}_{2}((A_{u}(g(\beta _{j}(x)),S^{2}))
&=&\lambda _{j}\mathcal{L}_{2}(S^{2})\mathcal{M}_{0}(g^{-1}(u,\infty )) \\
&=&4\pi \left\{ \frac{\sum_{\ell }b^{2}(\frac{\ell }{B^{j}})\frac{2\ell +1}{%
4\pi }C_{\ell }\frac{\ell (\ell +1)}{2}}{\sum_{\ell }b^{2}(\frac{\ell }{B^{j}%
})\frac{2\ell +1}{4\pi }C_{\ell }}\right\} 2(1-\Phi (\sqrt{u}))
\end{eqnarray*}%
implying that%
\[
\mathbb{E}\mathcal{L}_{2}((A_{u}(g(\beta _{j}(x)),S^{2}))=4\pi \times
2(1-\Phi (\sqrt{u}))\text{ .}
\]

\subsection{The Cubic Case $g(x)=x^{3}$}

Again by applying \eqref{GKF2}, we obtain for needlet components%
\begin{eqnarray*}
\mathbb{E}\mathcal{L}_{0}(A_{u}(\beta _{j}^{3}(x);S^{2}))
&=&\sum_{k=0}^{2}(2\pi )^{-k/2}\mathcal{L}_{k}(S^{2})\mathcal{M}_{k}^{%
\mathcal{N}}((\sqrt[3]{u},\infty )) \\
&=&2(1-\Phi (\sqrt[3]{u}))+2\frac{\sum_{\ell }b^{2}(\frac{\ell }{B^{j}})%
\frac{2\ell +1}{4\pi }C_{\ell }\frac{\ell (\ell +1)}{2}}{\sum_{\ell }b^{2}(%
\frac{\ell }{B^{j}})\frac{2\ell +1}{4\pi }C_{\ell }}\frac{e^{-(\sqrt[3]{u}%
)^{2}/2}}{\sqrt{2\pi }}\sqrt[3]{u}\text{ ,}
\end{eqnarray*}%
and likewise%
\[
\mathbb{E}\mathcal{L}_{1}(A_{u}(\beta _{j}^{3}(x);S^{2}))=2\pi \left\{ \frac{%
\sum_{\ell }b^{2}(\frac{\ell }{B^{j}})\frac{2\ell +1}{4\pi }C_{\ell }\frac{%
\ell (\ell +1)}{2}}{\sum_{\ell }b^{2}(\frac{\ell }{B^{j}})\frac{2\ell +1}{%
4\pi }C_{\ell }}\right\} ^{1/2}e^{-(\sqrt[3]{u})^{2}/2}.
\]%
Finally%
\[
\mathbb{E}\mathcal{L}_{2}(A_{u}(\beta _{j}^{3}(x);S^{2}))=4\pi (1-\Phi (\sqrt%
[3]{u}))\text{ .}
\]

\end{document}